\documentclass{icrc29}
\usepackage{graphicx,amssymb,amsmath,times}
\begin{document}
\newenvironment{my_enumerate}
{\begin{enumerate}
    \setlength{\topsep}{1pt}
    \setlength{\partopsep}{0pt}
    \setlength{\itemsep}{0pt}
    \setlength{\parskip}{0pt}
    \setlength{\parsep}{0pt}
    }
{\end{enumerate}}
%
%
%Title of paper
\title[Studies of Air Showers above $10^{18}$ eV with the CHICOS Array]
{Studies of Air Showers above $10^{18}$ eV with the CHICOS Array}
\author[T.W. Lynn et al.]
{T.W. Lynn$^a$, E. Brobeck$^a$, B.E. Carlson$^a$, C.J. Jillings$^a$, M.B. Larson$^a$, R.D. McKeown$^a$,
\newauthor
J.E. Hill$^b$, B.J. Falkowski$^c$, R. Seki$^c$, J. Sepikas$^d$, G.B.
Yodh$^e$, D. Wells$^f$, K.C. Chan$^g$.
\\
        (a) W.K. Kellogg Radiation Laboratory, California Institute of Technology,
        Pasadena, CA  91125, USA.
          \\
        (b) Department of Physics, California State University at Dominguez Hills,
        Carson, CA  90747, USA.
        \\
        (c) Department of Physics and Astronomy, California State
        University at Northridge, Northridge, CA  91330, USA.
        \\
        (d) Department of Astronomy, Pasadena City College,
        Pasadena, CA 91106, USA.
        \\
        (e) Department of Physics and Astronomy, University of California at Irvine,
        Irvine, CA  92697-4575, USA.
        \\
        (f) Department of Physics and Astronomy,
        California State University at Los Angeles, Los Angeles, CA  90032, USA.
        \\
        (g) The Chinese University of Hong Kong, Shatin, N.T., Hong Kong SAR.
        }
\presenter{Presenter: T. W. Lynn (theresa@caltech.edu),
usa-lynn-T-abs1-he14-oral}

\maketitle

\begin{abstract}

{\sc CHICOS} (California HIgh school Cosmic ray ObServatory) is
presently an array of more than 140 detectors distributed over a
large area ($\sim$400 km$^2$) of southern California, and will
consist of 180 detectors at 90 locations in the near future. These
sites, located at area schools, are equipped with computerized
data acquisition and automatic nightly data transfer (via
internet) to our Caltech lab.  The installed sites make up the
largest currently operating ground array for ultra-high energy
cosmic ray research in the northern hemisphere.  The goal of
CHICOS is to provide data related to the flux and distribution of
arrival directions for ultra-high energy cosmic rays.

\vspace{-0.1in} We have performed detailed Monte-Carlo
calculations to determine the density and arrival-time
distribution of charged particles in extensive air showers for the
CHICOS array. Calculations were performed for proton primaries
with energies $10^{18}$ to $10^{21}$ eV and zenith angles out to
$\simeq 50^{\circ}$. We have developed novel parameterizations for
both distributions as functions of distance from the shower axis,
primary energy, and incident zenith angle. These parameterizations
are used in aperture calculations and reconstruction of shower
data, enabling preliminary analysis of ultra-high energy shower
data from CHICOS.
\end{abstract}

\section{Introduction}
\begin{figure}
\begin{center}
$\begin{array}{c@{\hspace{0.4in}}c}
\includegraphics[width=0.45\textwidth]{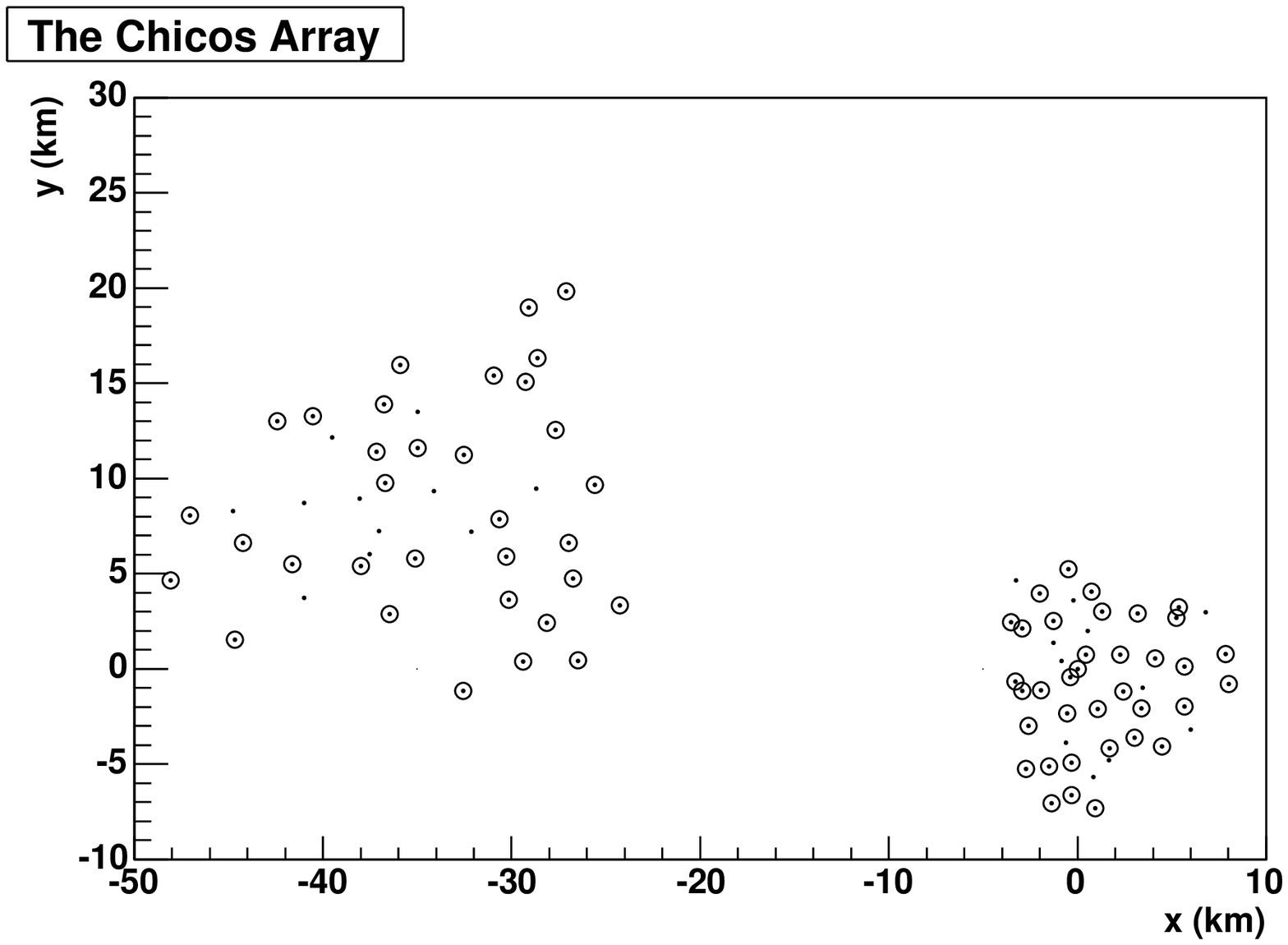} &
\includegraphics[width=0.45\textwidth]{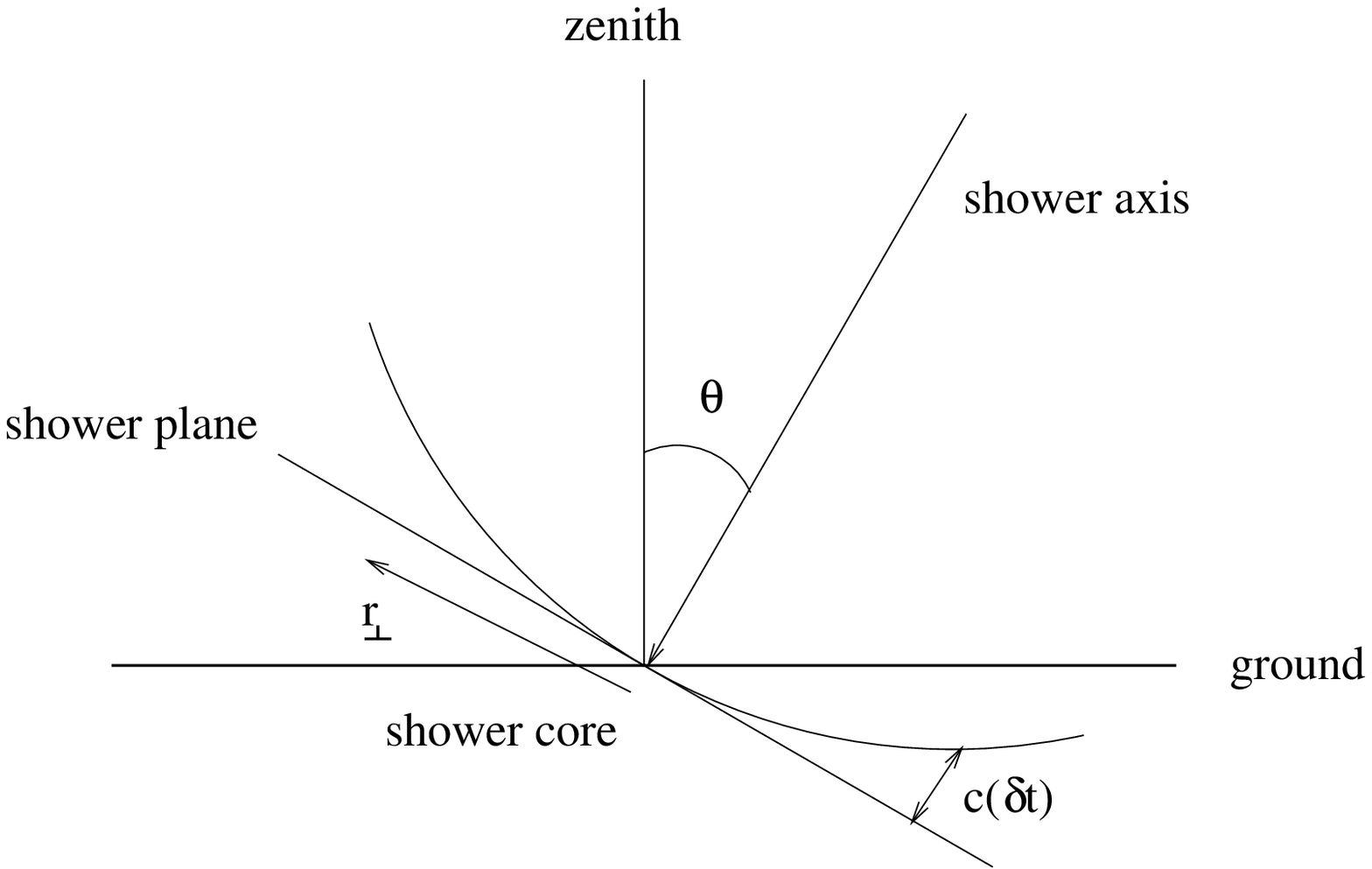}
     \\ [0.4cm]
\mbox{\bf (a)} & \mbox{\bf (b)}
\end{array}$
\end{center}
\caption{(a)The CHICOS array in the San Gabriel and San Fernando
Valleys of Los Angeles County, California. Each site contains two
scintillator detectors separated by $\sim$5m. Circled dots and
bare dots are sites installed and in progress, respectively. The
array averages approximately 250m above sea level, with individual
detector sites ranging from $\sim$100m to $\sim$400m in altitude.
 (b)The geometry of an extended air shower showing the definition of
$r_\perp$ and the time delay $\delta t$.\label{fig:arrayshower} }
\end{figure}

%Considerable uncertainty currently surrounds the flux of
%ultra-high energy cosmic rays (UHECRs) in the energy range above
%$10^{18}$ eV. UHECRs in this energy range have been observed in
%the northern hemisphere by the AGASA ground array in
%Japan\cite{agasa} and by the HiRes collaboration in
%Utah\cite{hires}.  The Pierre Auger Observatory is taking data and
%expanding in Mendoza Province, Argentina with both a ground array
%and air-fluoresence telescope\cite{auger}.  For primary particles
%in this energy range, the astronomical source or sources are
%unknown. Furthermore, for proton primaries, theory predicts a
%sharp dropoff in UHECR flux above E~10$^{19.5}$ eV, known as the
%GZK cutoff, due to photopion production on the cosmic microwave
%background\cite{gzk}. The observation of this cutoff is at present
%somewhat inconclusive and raises doubt as to our understanding of
%the propagation effects, primary particle identification, or other
%physics relevant at these energies.
%
{\sc CHICOS}, the California HIgh school Cosmic ray ObServatory,
is a large ground array of plastic scintillator detectors located
at schools and universities in the Los Angeles, California area.
Siting the array in a large urban area takes advantage of existing
infrastructure to power the array as well as school internet
connections for data transfer. The array has been observing
extensive air showers since early 2003, with the goal of measuring
the flux and arrival direction distribution for ultra-high energy
cosmic rays (above $10^{18}$eV in primary energy).
The CHICOS array covers an area of $\sim$400 km$^2$ with an
average site-to-site spacing of approximately 2 km (see Figure
\ref{fig:arrayshower}). {\sc CHICOS} uses thin ($\lesssim$ 10 cm)
plastic scintillators of area $\approx 1m^2$ to measure the energy
deposited in the detectors by charged particles. When a series of
individual detector hits passes cuts and is considered a candidate
shower event, the measured densities are used to determine the
location and energy of the primary cosmic ray. The time of hits in
each detector is used to determine the primary direction. To
properly reconstruct the arrival direction, it is necessary to
know the shape of the shower front. Furthermore, the direction and
energy are correlated since a more inclined shower must traverse a
greater atmospheric slant depth resulting in an ``older'' shower.
Since CHICOS is sited at an atmospheric depth of
$\sim$975g/cm$^2$, well beyond shower maximum, inclined showers
are attenuated relative to vertical showers. In the analyses
presented here, ``ground'' refers to an elevation of 250m MSL (see
Figure \ref{fig:arrayshower}a).

In order to reconstruct ultra-high energy showers from CHICOS
data, we have performed extensive simulations of proton-initiated
showers in the energy range $10^{18}$ to $10^{21}$ eV, using the
AIRES simulation package for extensive air showers\cite{aires}. We
present CHICOS-appropriate parameterizations for both the charge
density, or lateral distribution function (LDF), and the time
distribution function (TDF) of ground particles in an ultra-high
energy shower. These parameterizations are employed in the
analysis of CHICOS shower data using a chi-squared minimization
algorithm, giving preliminary reconstructions of energy and
arrival direction for observed events above $10^{18}$ eV.

\section{Shower Characteristics and Monte-Carlo Simulation}

Showers in the CHICOS energy range typically contain $10^9$ to
$10^{10}$ particles, and may have an density of $\geq$1/m$^2$ out
to 1km or more from the shower core. The time distribution of a
typical shower, at km-scale distance from the core, is
characterized by both shower curvature delay and shower time
spread in the range of several microseconds.  See Figure
\ref{fig:arrayshower}b for a schematic shower front and the
definition of terms used to describe the shower's geometry.

%{\sc Chicos} records the measured amplitudes and times of
%scintillator signals throughout the array. With a time constant of
%$\sim$90ns, each detector is expected to record individual shower
%particles as distinct hits.
The CHICOS data acquisition system samples the amplitudes and
times of scintillator signals at 12.5ns intervals. Multiple hits
within a shower are individually resolved.  Reconstructing primary
energy and direction from this data requires mathematical models
for both the LDF, $\rho(r_\perp;E,\theta)$, and the arrival time
distribution, $T_d(\delta t;r_\perp,E,\theta)$.

We used version 2.6.0 of the AIRES extensive air-shower simulation
program\cite{aires} to simulate proton-initiated UHECR showers.
Hadronic interactions were modeled with QGSJET\cite{qgsjet}. We
performed simulations at zenith angles $\cos\theta = (0.95, 0.85,
0.75,0.65)$and primary energies $log_{10}(E[eV]) =
(18,18.5,19,19.5,20,20.5)$.
Simulations at these energies almost always employ {\emph
{thinning}} or statistical sampling of particles. We set thinning
to begin at $E_{th} = 10^{-7} \times E_{primary}$, with the AIRES
statistical weight factor set to $W_{f}^{(EM)}=1$. Shower
particles were tracked down to $E_{e^\pm,\gamma}=1MeV$ and
$E_{\mu^\pm}=20MeV$. Finally, all ground particles were required
to have energy $\geq 5MeV$ to correspond to CHICOS detector
thresholds. We performed 10 simulation runs at each set of shower
parameters.
%AIRES also imposes a
%cutoff energy, or energy below which particles are no longer
%tracked in the simulation. We used a cutoff energy of X for
%electrons, positrons and gamma rays, and a cutoff of Y for muons.
%In the plots and fits shown below, an overall cutoff energy of
%5MeV has been applied after the simulation to more closely reflect
%the CHICOS detector threshold.
%
%We performed 10 simulation runs at each set of shower parameters.
%For each run we obtained a list of all particles that reached sea
%level (0 m MSL) as well as all the particles that crossed surfaces
%250m and 500 m above sea level. For the analyses shown here we use
%the particles at 250 m. The data files were approximately 200 to
%500 Mbytes large.
%
In the analysis four categories of histograms were considered:
\vspace{-0.15in}
\begin{my_enumerate}
  \item The number of electrons and positrons as a function of
$r_{\perp}$ in 50-m bins from 100m to 10000m.
  \item The number of electrons and positrons as a function of
  $r_{\perp}$ and $\delta t$ in 50m and 50ns bins respectively.
  \item The same (items 1 and 2) for muons ($\mu^+$ and $\mu^-$).
\end{my_enumerate}
\vspace{-0.15in}
These histograms were averaged over the ten runs
and the standard deviation of the runs was used as the uncertainty
in the histogram.

\section{Charge Density and Time Distributions}
%\section{Results and Conclusions}

To model the LDF, we began with the approach of the AGASA
experiment, which was similar to CHICOS in many ways but with
important differences in location and hardware.  AGASA used a
parameterization of a modified NKG formula\cite{agasa-ldf}.
%The AGASA formula, used in analysis of their results,
%is\cite{agasa_ldf}:
%\begin{equation}
%\rho (r_{\perp}) \propto \left( \frac{r}{R^{\sc AGASA}_M}\right)
%^{-1.2} \left( 1+\frac{r}{R^{\sc{AGASA}}_M}\right)^{-\eta+1.2}
%\left\{ 1+ \left(\frac{r}{1000}\right)^2 \right\} ^{-0.6}
%\end{equation}
%where $R^{AGASA}_M = 91.6m$ is the Moli\`ere radius at the
%altitude of AGASA and $\eta=(3.97\pm 0.13) - (1.79\pm
%0.62)(\sec\theta -1)$.
Thus we also began with the functional form:
\begin{equation}
\rho (r_{\perp}) = {\cal C}\left( \frac{r_{\perp}}{R_M}\right)
^{-\alpha} \left( 1+\frac{r_{\perp}}{R_M}\right)^{-\eta+\alpha}
\left\{ 1+ \left(\frac{r_{\perp}}{1000}\right)^2 \right\}
^{\delta} \label{eq:ldf}
\end{equation}
where $r_{\perp}$ is in meters and $\rho$ is in particles/$m^2$.
All parameters are potentially functions of the zenith angle
$\theta$, energy $E$ (in eV), and species of the primary cosmic
ray. Note that $R_M$ is an {\emph {effective}} Moli\`ere radius
allowed to vary in order to fit the shower shape subject to the
energy threshold.

\begin{figure}
\begin{center}
$\begin{array}{c@{\hspace{0.4in}}c}
\includegraphics[width=0.45\textwidth]{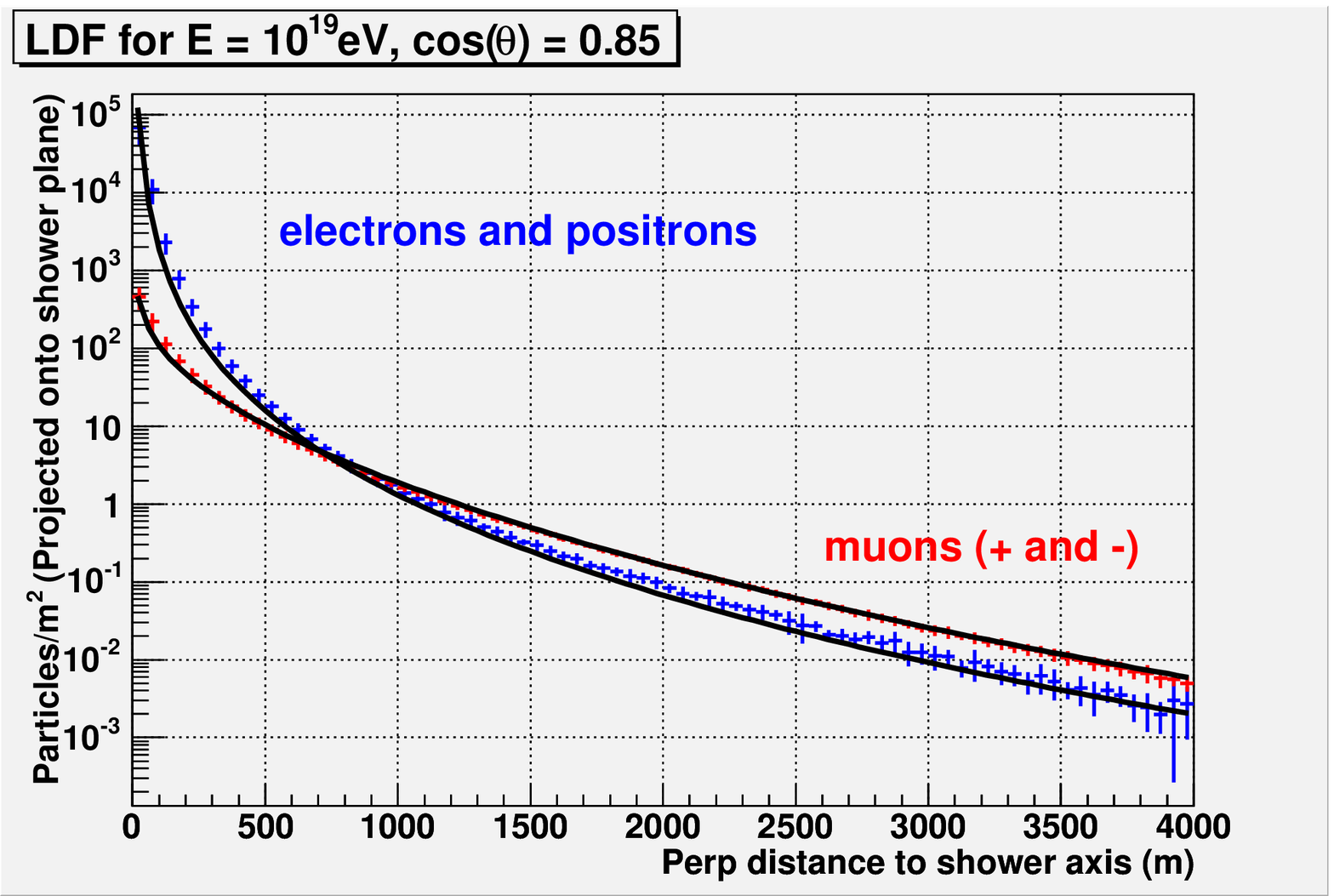} &
\includegraphics[width=0.45\textwidth]{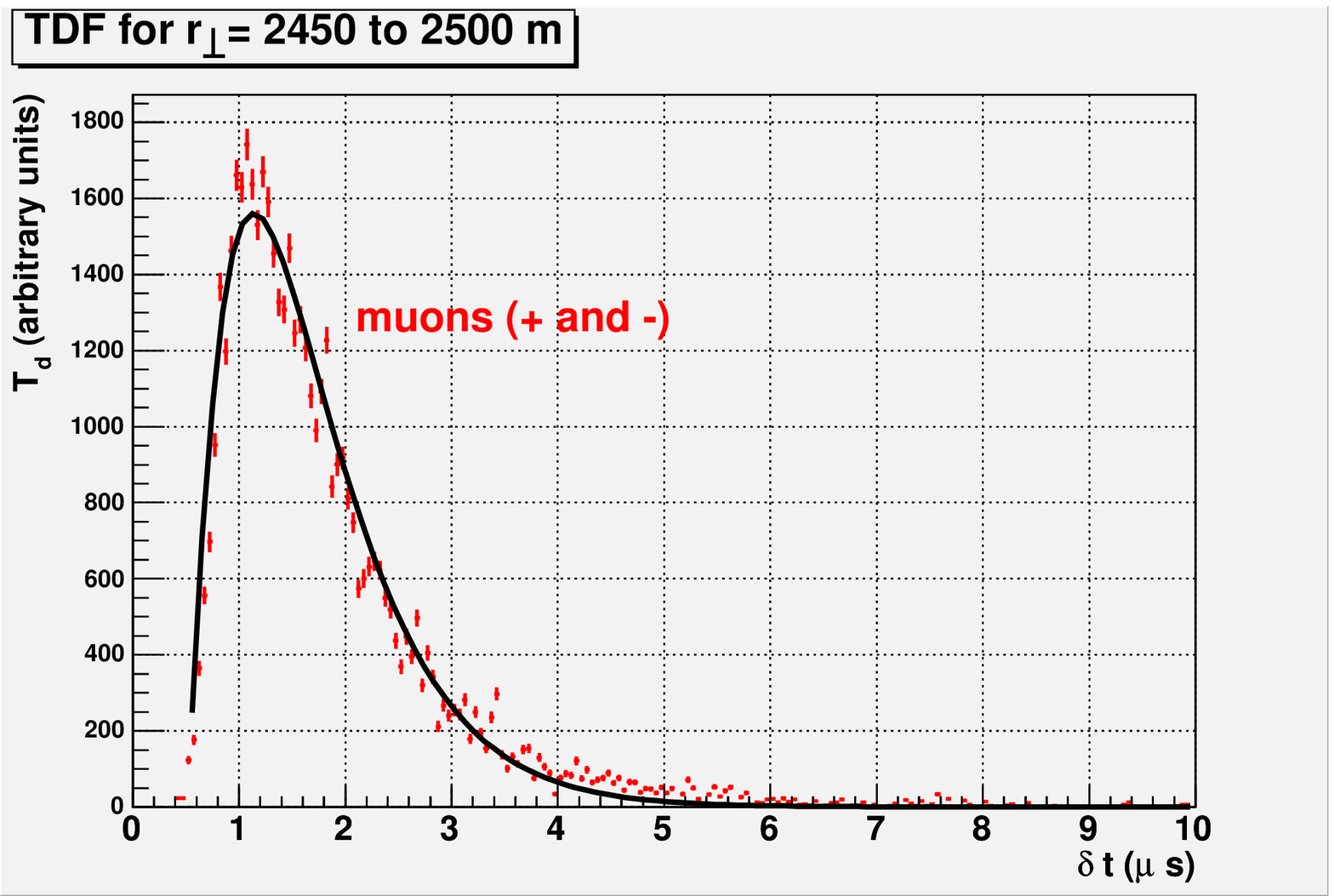}
     \\ [0.4cm]
\mbox{\bf (a)} & \mbox{\bf (b)}
\end{array}$
\end{center}
\caption{(a) The behavior of particle densities as a function of
$r_\perp$ for proton primary of energy $10^{19}\,\mbox{eV}$ and
zenith angle of $\cos\theta = 0.85$. Points with error bars are
AIRES output (mean and standard deviation of 10 runs).  The solid
curve overlay shows the final muon and electron LDF
parameterizations. (b)Arrival time distribution for muons at radius 
$2450m < r_{\perp} <2500m$, shower parameters of (a).
\label{fig:samples} }
\end{figure}

Figure~\ref{fig:samples} shows the simulation with
$E_{\mbox{{\small primary}}} = 10^{19} \mbox{eV}$ and $\cos\theta
= 0.85$.  This example shows the typical contrast between muon and
electron distributions, motivating us to parameterize the
distributions independently of one another.  For each species, we
employed an iterative process of fitting all individual
histograms, fixing or partially fixing parameter values, and
refitting for the remaining parameters.  Our final LDF
parameterizations are given by Equation \ref{eq:ldf} with the
parameters listed in Table \ref{table:param}.  Examples of these
formulae are drawn as solid curves over the AIRES output
histograms in Figure \ref{fig:samples}a.

\begin{table}
\caption{\label{table:param} LDF Parameters}
\begin{center}
\begin{tabular}{||c||c|c|c|c|c||} \hline \hline

 Particle Type  & $R_M$ & $\alpha$ & $\delta$ & $\eta$ & $log_{10}(C)$\\

                             \hline \hline

Electron        &  $2477$  & $2.513$ & $0.03107$ & $8.391-5.315(sec(\theta)-1)$ & $0.1-1.45(sec(\theta)-1)$ \\
                & & & & &    $+0.96(log_{10}(E)-19)$\\
                \hline

Muon        &  $2560$  &  $0.7701$ & $0.01939$ & $9.020-2.552(sec(\theta)-1)$ & $1.2-0.72(sec(\theta)-1)$ \\
            & & & & & $+0.97(log_{10}(E)-19)$ \\
                             \hline \hline
\end{tabular}
\end{center}
\end{table}

Turning to the shower front curvature delay and the spread of
arrival times, we note that there is no well-motivated and
accepted model similar to the modified NKG formula. Therefore a
purely phenomenological model was developed. As with the lateral
distribution function, muons and electrons are treated separately
in our parameterizations.  Rather than fitting an average arrival
time and spread, we model the shape of the shower with a full time
distribution function for curvature time delay:
\begin{equation}
T_d(\delta t;r_\perp,E,\theta)  =
        \left\{\begin{array}{ll}
        N(\delta t-a)^b\exp[-c(\delta t-a)]&  \mbox{if  $\delta t\geq a$};
        \\
        0 &    \mbox{if  $\delta t < a$}.\end{array} \right. \label{eq:tdf}
\end{equation}
where $N$ is an overall normalization constant. We find that in
fact, over the CHICOS energy range, $a$, $b$, and $c$ may be
considered functions of $r_\perp$ and $\theta$ without introducing
a dependence on $E$. An example of the muon TDF model is shown in
Figure \ref{fig:samples}b.  We are presently working to integrate
the final TDF parameterization in our shower reconstruction
algorithm; it will replace a modified AGASA time delay formula
currently in use.

\section{Conclusions}

The AIRES simulations and resulting shower parameterizations
presented here are appropriate for the CHICOS location and
instrumentation.  As such, these results allow preliminary shower
reconstruction; final reconstructed energies and directions await
further refinements in our treatment of individual detector
response. A smaller-scale ``Chiquita'' array has been deployed
within CHICOS to build a data set for showers at energies $\sim
10^{16}$ to $10^{18}$eV; analysis of the Chiquita data is
presented elsewhere in these proceedings and provides us with a
test bed for the reconstruction technique and detector
modeling\cite{chiquita-icrc}. In analyses of Chiquita showers we
have also performed limited comparisons of AIRES outputs with the
results of the CORSIKA simulation package\cite{corsika}, finding
good agreement between the two.

\section{Acknowledgements}

We thank S. Ligocki for recent work on shower analysis software.
We thank Los Alamos National Laboratory for donation of
scintillator detectors, and IBM for donation of computer
equipment.  We acknowledge support from the National Science
Foundation, the Weingart Foundation, and the California Institute
of Technology.

\end{document}